# Aspects Regarding Operations
# with Fuzzy Processes

**Lucian Luca, Lucian L. Luca,**
**"Tibiscus" University of Timişoara**

**ABSTRACT.** This paper introduces the notion of *fuzzy process* as a formalism for the idea of fuzzy contact between a device and its environment. The notions of absolute correctness and relative correctness are defined. In order to work with concurrency it has been built an approach to manipulate the interactive processes as a single process and the resulted behavior has been observed.

## Introduction

We introduce the concept of fuzzy process, formalism for the idea of fuzzy contract between a device and its environment. Such a contract specifies the device-environment interface in terms of executions.

The executions can be sequences of events, time functions, etc.; however we consider them only as elements of an arbitrary set $E$. Since we do not speculate about the structure of the elements of $E$, we can endow the processes with algebraic properties and we can demonstrate some of their characteristics.

There are many meanings of the term process. First, a process is related to flows in the real world, to progressive observable changes of the structure of a system. Then, a process is a structured change, namely there is a pattern of events which an observer can recognize among the occurrences of the process.

Another important distinction is between the natural and artificial processes. We know that artificial processes are built by people and they exist with the assumed purpose to change the condition of the real world or

205

to restrain it so that it may meet human requirements. Artificial processes have to be started by an agent to transform the world, while the natural processes do not.

If the agent is a machine, there is predictability and determinism in the process, allowing a precise description. If the agent is human, there is no guarantee that the anticipated event will take place; it is indeterminist.

On the other hand, the human agents have an adaptive capacity: a person can perceive a mistake or an error in the definition of a process and he can take the initiative to correct it. Thus, the objective can be achieved even if it was originally wrong, and this is the essential difference between the world of cars and people.

Further on, by interacting systems we understand systems that can be coupled and compared. The space of the fuzzy processes is just a unified theory of interacting systems, including concurrent systems, as a particular case.

To simplify the model, we will use roles, namely a set of standards, descriptions and rules assigned to a person or a position. The two main roles that we use are the device and its environment. Let us note that a role has two aspects: responsibility (rights, powers, duties) and a template for actions, some of them involving the interaction with other roles.

Our work is a generalization of [Neg95], [Neg98]. The basic ideas of this author are: to describe interactive systems, using the notion of abstract and primitive execution. A process is a pair of two sets of executions: one for the device and another for the environment. We develop the model, adding the fuzziness.

## 1 The basic formalism

Let $E$ be the set of all the executions as $\Delta : E \to [0,1]$ and $\Gamma : E \to [0,1]$ two fuzzy sub-sets of $E$. In what follows, we note with:

$$X = \{x \in E \mid \Delta(x) > 0\}, \ Y = \{x \in E \mid \Gamma(x) > 0\}, \ B = \{x \in E \mid \Delta(x) = \Gamma(x) = 0\}$$

and respectively we call:
 X – the set of accessible execution;
 Y – the set of acceptable execution;

B – the set of violations.

Moreover, we note $\Delta_X = \Delta_{/X}$, $\Gamma_Y = \Gamma_{/Y}$.

**Definition 1**: *The pair $p = (\Delta_X, \Gamma_Y)$, where $\Delta_X$ and $\Gamma_Y$ are defined as above, is called **fuzzy process** over E.*

*The set of all fuzzy processes over a pair of crisp sub-sets X and Y of E, as above, is called **the space of fuzzy processes of (X, Y)**, and the set of all the fuzzy process over E is called **the space of fuzzy processes of E**.* ∎

A fuzzy process $p = (\Delta_X, \Gamma_Y)$ is a contract between the device and its environment: the device can ensure that only executions of *X* may occur, while the environment guarantees that only the executions of *Y* may occur.

The device can access, respectively accept, an execution. An execution $x \in E$ is *(X,Y)-completely accesibile* if $\Delta(x) = 1$ and it is *(X,Y)-completely acceptabile* if $\Gamma(x) = 1$.

According to the classic (crisp) theory of the sets, a fuzzy process partitions the set *E* of all the executions in four disjunctive sub-sets $\widetilde{X}, \widetilde{Y}, X \cap Y, B$ where:

$$\widetilde{X} = \{x \in E \mid \Delta(x) = 0 \wedge \Gamma(x) > 0\}, \; \widetilde{Y} = \{x \in E \mid \Gamma(x) = 0 \wedge \Delta(x) > 0\}.$$

Obviously, $\widetilde{X} \cap \widetilde{Y} = \phi$

We call the elements of $\widetilde{X}$ *escapes* and they should be avoided by the device. We call the elements of $\widetilde{Y}$ *rejects* and they should be avoided by the environment. Together, the elements of $\widetilde{X} \cup \widetilde{Y}$ are called **violations.**

The agreement states that only the executions of $X \cap Y$ are allowed to appear in the presence of the device. For this reason, $X \cap Y$ is also called **the contract set**, and the executions $x \in X \cap Y$ are also called **goals,** because they are legal both for the device, and for the environment. The set X contains many executions for which the device respects the agreement (and the environment may respect it or not). The set Y contains the executions for which the device respects the agreement (while the device nay respect it or not). In the set $X \cup Y$ we still can have many executions with the degree of acceptability or the accessibility equally with zero, but not both zero for a given execution (Figure 1).

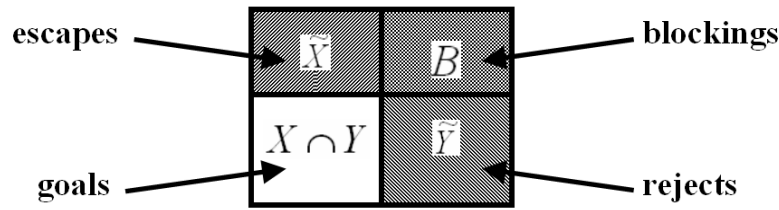

*Figure 1: Types of executions*

## 2 Corectness conditions

The property of ***absolute corectness*** in the space of fuzzy processes formalizes the fact that a device operates correctly through itself, meaning that the device does not impose any requirement to the environment. In this respect, the device is called ***robust***. In terms of the avoided executions, this property leads to a set of rejects, void for the corresponding process.

The symmetric property is that the device does not offer any guarantee to the environment. In terms of avoided executions, this leads to a set of escapes void for the corresponding process. In this respect, the device is called ***chaotic.***

**Definition 2:** *The fuzzy process* $p = (\Delta_X, 1_E)$ *is called* ***robust*** *and the fuzzy process* $p = (1_E, \Gamma_Y)$ *is called* ***chaotic***. ∎

In what follows, we note with $R_E$ and $H_E$ the set of robust processes and respectively chaotic over E. The process $p = (1_E, 1_E)$ is the only fuzzy process that is robust and chaotic at the same time.

**Definition 3:** *The void fuzzy process is given by* $\Omega = (1_E, 1_E)$. ∎

This process has no escapes or rejection. Thus, it offers no guarantees and no restrictions on the environment.

The property of ***relative correctness*** in the space of the fuzzy processes formalizes the fact that a fuzzy process $q$ is a satisfactory substitute for a fuzzy process $p$. $q$ should impose fewer requirements on the environment and offer more guarantees than $p$. In terms of avoided executions, it means that $q$ has a bigger set of acceptable executions and a smaller set of available executions smaller comparing to those corresponding to $p$.

**Definition 4:** *Let us have two fuzzy processes for the same set of executions* E, $p = (\Delta^p_{X_p}, \Gamma^p_{Y_p})$ *and.* $q = (\Delta^q_{X_q}, \Gamma^q_{Y_q})$ *We say that the fuzzy process p is refined by the fuzzy process q and we write* $p \sqsubseteq q$ *if and only if* $(\Delta^p(x) \geq \Delta^q(x)) \wedge (\Gamma^p(x) \leq \Gamma^q(x)), \forall x \in E.$ ∎

Several interesting properties related to the refining can be found in [Luc03].

## 3 Operations with fuzzy processes

To work with the concurrency, we build an approach to manipulate the interactive processes as a single process and we consider their resulted behavior. Thus, we have to consider both types of interactions, between the devices, as well as between the environments. Therefore, we define two operations that give us the general behavior: the product will be seen as the law of composition of the devices and the sum will be seen as the law of composition of the environment.

Since we are not interested in the interactions between the fuzzy processes for which the set $B = \phi$ (Blockings), in what follows we consider only fuzzy processes for which there are no blocks; thus a fuzzy process being given $p = (\Delta^p_{X_p}, \Gamma^p_{Y_p})$, we consider $\{x \in E \mid \Delta(x) = \Gamma(x) = 0\} = \phi$.

We can consider anytime such processes, taking $E' = E - B$, that is we consider only those executions for which $\Delta^p(x) > 0$ or $\Gamma^p(x) > 0$.

**Definition 5:** *Let us have two fuzzy processes over the same set* E *of executions,* $p = (\Delta^p_{X_p}, \Gamma^p_{Y_p})$ *and* $q = (\Delta^q_{X_q}, \Gamma^q_{Y_q})$ *the **product** of the p and q processes is the process* $p \otimes q$ *so that:*

$$X_{p \otimes q} = X_p \cap X_q$$

$$Y_{p \otimes q} = (Y_p \cap Y_q) \cup (\tilde{X}_p \cap \tilde{Y}_q) \cup (\tilde{Y}_p \cap \tilde{X}_q)$$

$$\Delta^{p \otimes q}(x) = \begin{cases} \min\limits_{x \in X_p \cap X_q} \{\Delta^p_{X_p}(x), \Delta^q_{X_q}(x)\} > 0 \\ 0, \quad \hat{i}n \; rest \end{cases}$$

$$\Gamma^{p \otimes q}(x) = \begin{cases} \min\limits_{x \in Y_p \cap Y_q} \{\Gamma^p_{Y_p}(x), \Gamma^q_{Y_q}(x)\} > 0 \\ \min\limits_{x \in \tilde{X}_p \cap \tilde{Y}_q} \{\Gamma^p_{Y_p}(x), \Delta^q_{X_q}(x)\} > 0 \\ \min\limits_{x \in Y_p \cap \tilde{X}_q} \{\Gamma^q_{Y_q}(x), \Delta^p_{X_p}(x),\} > 0 \\ 0, \quad \hat{i}n \; rest \end{cases} \quad \blacksquare$$

Analogically, we define the sum operation:

**Definition 6:** *Let us have two fuzzy processes over the same set E of executions,* $p = (\Delta^p_{X_p}, \Gamma^p_{Y_p})$ *and* $q = (\Delta^q_{X_q}, \Gamma^q_{Y_q})$ *,* **the sum** *of the processes p şi q is the process* $p \oplus q$ *so that:*

$$X_{p \oplus q} = (X_p \cap X_q) \cup (\tilde{X}_p \cap \tilde{Y}_q) \cup (\tilde{Y}_p \cap \tilde{X}_q)$$

$$Y_{p \oplus q} = Y_p \cap Y_q$$

$$\Delta^{p \oplus q}(x) = \begin{cases} \min\limits_{x \in X_p \cap X_q} \{\Delta^p_{X_p}(x), \Delta^q_{X_q}(x)\} > 0 \\ \min\limits_{x \in \tilde{X}_p \cap \tilde{Y}_q} \{\Gamma^p_{Y_p}(x), \Delta^q_{X_q}(x)\} > 0 \\ \min\limits_{x \in \tilde{Y}_p \cap \tilde{X}_q} \{\Gamma^q_{Y_q}(x), \Delta^p_{X_p}(x),\} > 0 \\ 0, \quad \hat{i}n \; rest \end{cases}$$

$$\Gamma^{p \oplus q}(x) = \begin{cases} \min\limits_{x \in Y_p \cap Y_q} \{\Gamma^p_{Y_p}(x), \Gamma^q_{Y_q}(x)\} > 0 \\ 0, \quad \hat{i}n \; rest \end{cases}$$

The product or sum operations are indempotent, commutative, associative and they admit the void process as a identity element. These properties are immediate from the definitions of the operations with classic sets.

**Preposition 1:** *Let us have the fuzzy processes p, q and r over the same set E of executions:*

| | | | | |
|---|---|---|---|---|
| *i)* | $p \otimes p = p$ | | *i')* | $p \oplus p = p$ |
| *ii)* | $p \otimes (q \otimes r) = (p \otimes q) \otimes r$ | | *ii')* | $p \oplus (q \oplus r) = (p \oplus q) \oplus r$ ∎ |
| *iii)* | $p \otimes q = q \otimes p$ | | *iii')* | $p \oplus q = q \oplus p$ |
| *iv)* | $p \otimes \Omega = p$ | | *iv')* | $p \oplus \Omega = p$ |

Informally, iv) and iv' show that the introduction of a void system does not alter the behavior of the system.

The relationship resulting from definition 4 introduces a new unar operator on the set of all fuzzy processes over an *E* set of executions: the **reflection** of a fuzzy process *p*. Informally, if *p* is an agreement between a device and its environment, from the point of view of the device, then reflection of p represents the proper understanding from the point of view of the environment.

**Definition 7**: The **reflection** of a fuzzy process $p = (\Delta_X, \Gamma_Y)$ *is a fuzzy process* $-p = (\Gamma_Y, \Delta_X)$ ∎

The reflection is its own inverse, it reverses the refining and commutes $R_E$ and $H_E$ between them.

**Proposition 2**: *Let us have two fuzzy processes p and q over the same set of executions E,*

$$i) \quad -p = -- p$$

$$ii) \quad p \sqsubseteq q \quad \Leftrightarrow \quad -q \sqsubseteq -p$$

$$iii) \quad p \in R_E \quad \Leftrightarrow \quad -p \in H_E$$

*The demonstration is immediate from the definitions of* $-, \sqsubseteq, R_E$ *şi* $H_E$ ∎

There are situations when we do not have complete information about the behavior of a particular device or environment. They may be viewed as choices between alternatives of behavior and they can be modeled employing two new operations over the fuzzy processes: one for choices between the devices and another for choices between environments.

**Definition 8:** L*et us have two fuzzy processes over the same set of executions E,* $p = (\Delta^p_{X_p}, \Gamma^p_{Y_p})$ *and* $q = (\Delta^q_{X_q}, \Gamma^q_{Y_q})$ ,

their ***meet,*** p ⊓ q *is:*

$$p \sqcap q = (\Delta^{p \sqcap q}, \Gamma^{p \sqcap q}), \text{ where}$$

$$\Delta^{p \sqcap q}(x) = \max_{x \in X_p \cap X_q} \{\Delta^p_{X_p}(x), \Delta^q_{X_q}(x)\}$$

$$\Gamma^{p \sqcap q}(x) = \min_{x \in Y_p \cap Y_q} \{\Gamma^p_{Y_p}(x), \Gamma^q_{Y_q}(x)\}$$

*and their* ***join***, $p \sqcup q$ , *is:*

$$p \sqcup q = (\Delta^{p \sqcup q}, \Gamma^{p \sqcup q})$$

$$\Delta^{p \sqcup q}(x) = \min_{x \in X_p \cap X_q} \{\Delta^p_{X_p}(x), \Delta^q_{X_q}(x)\}$$

$$\Gamma^{p \sqcup q}(x) = \max_{x \in Y_p \cap Y_q} \{\Gamma^p_{Y_p}(x), \Gamma^q_{Y_q}(x)\} \quad \blacksquare$$

Let us have two alternative fuzzy processes, their intersection and join fuzzy processes can behave as any of the alternatives and this behavior choice is made for each execution.

*The intersection* of fuzzy processes models an option between devices. If an execution is accessible for any of the alternative devices, the *intersection device* will not guarantee the avoidance of this execution, because we do not know what device was chosen. If an execution is a rejection for one of the alternative devices, the *intersection device* will also require from its environment to avoid that execution in order to obstruct the possibility for a device to reject that execution.

*The intersection* can be seen as a situation with a partner, in which we choose an execution and then the partner chooses a device, so that it should maximize the possibility of errors occurring in a system consisting of a device and its environment.

Dually, the join models a non-deterministic choice between environments. The device of $p \sqcup q$ has a set of acceptable executions large enough and a set of accessible executions small enough to adapt itself to an environment that may choose to behave either like –*p* or like –*q*.

From the definitions of the **top** fuzzy processes, $\top = (0_E, 1_E)$ and **bottom**, $\bot = (1_E, 0_E)$, it immediately follows the proposition:

**Proposition 3:** *Let us have a fuzzy process* $p = (\Delta_{X_p}^p, \Gamma_{Y_p}^p)$ *over the set E of executions*

| | |
|---|---|
| *i)* $\quad p \sqcap \top = p$ | *i')* $\quad p \sqcup \bot = p$ |
| *ii)* $\quad p \sqcup \top = \top$ | *ii')* $\quad p \sqcap \bot = \bot$ |
| *iii)* $p \otimes \top = \top$ | *iii')* $\quad p \oplus \bot = \bot$ |
| *iv)* $\quad - \top = \bot$ $\qquad$ ■ | |

The *top* process has the interesting property that, if it is inserted into a system, then all the possible defects of the system are eliminated. *iii)* indicates that the insertion of a *top* fuzzy process in an arbitrary system will be also *top,* that makes it robust.

The spaces of the fuzzy processes have symmetries between the environment and device.

**Proposition 4:** *Let us have two fuzzy processes over the same set of executions E,* $p = (\Delta_{X_p}^p, \Gamma_{Y_p}^p)$ *and* $q = (\Delta_{X_q}^q, \Gamma_{Y_q}^q)$

*i)* $\qquad -(p \otimes q) = -p \oplus -q$

*ii )* $\qquad -(p \oplus q) = -p \otimes -q$

*iii)* $\qquad -(p \sqcap q) = -p \sqcup -q$

*iv)* $\qquad -(p \sqcup q) = -p \sqcap -q$

*The demonstration is immediate from the definition of the operators over the system of the fuzzy processes.* ■

The propositions 2 and 4 show that the reflection is an isomorphism over the space of the fuzzy processes.

**References**

[LD01] L. Luca, I. Despi - *Toward a Definition of Fuzzy Processes*, Proceedings of the 5th International Symposium on Economics Informatics, Bucharest, pp. 855-859, May 2001.

[Luc03]    Luca, L. – *Spaţii de procese fuzzy*, Editura Mirton, Timişoara, 2003

[Neg95]    R. Negulescu - *Process spaces*, Technical Report CS-95-48, Department of Computer Science, University of Waterloo, Ontario, Canada, December, 1995.

[Neg98]    R Negulescu - *Process Spaces and Formal Verification of Asynchronous Circuits*, PhD thesis, Department of Computer Science, University of Waterloo, Ontario, Canada, August, 1998.